\documentclass[aip,jcp,reprint]{revtex4-1}
\usepackage{amsmath,amssymb,bm}
\usepackage{graphicx}
\usepackage[bookmarks=false,breaklinks=true]{hyperref}
\usepackage[utf8]{inputenc}
\usepackage{newunicodechar}
\usepackage{pgf}
\usepackage{siunitx}
\usepackage{xcolor}

\newunicodechar{−}{\ensuremath{-}}

\hypersetup{
  pdfstartview = FitH,
  pdftoolbar   = false,
  pdfmenubar   = true,
  colorlinks   = true,
  linkcolor    = blue!50!black,
  urlcolor     = blue!50!black,
  citecolor    = blue!50!black
}

\begin{document}

\title{Nanoconfined catalytic {\AA}ngstr{\"o}m-size motors}
\author{Peter~H.~Colberg}
\email{pcolberg@chem.utoronto.ca}
\author{Raymond~Kapral}
\email{rkapral@chem.utoronto.ca}
\affiliation{Chemical Physics Theory Group, Department of Chemistry,
University of Toronto, Toronto, Ontario M5S 3H6, Canada}

\begin{abstract}
Self-propelled chemically powered synthetic micron and nano-scale motors are
being intensively studied because of the wide range of potential applications
that exploit their directed motion.
This paper considers even smaller {\AA}ngstr{\"o}m-size synthetic motors.
Such very small motors in bulk solution display effects arising from their
self-propulsion.
Recent experiments have shown that small-molecule catalysts and single enzyme
molecules exhibit properties that have been attributed to their chemical
activity.
Molecular dynamics is used to investigate the properties of very small
{\AA}ngstr{\"o}m-size synthetic chemically powered sphere-dimer motors in a
simple atomic-like solvent confined between walls separated by distances of
tens of nanometers.
Evidence for strong structural ordering of the motors between the walls, which
reflects the finite size of solvent molecules and depends on solvent depletion
forces, is provided.
Dynamical properties, such as average motor velocity, orientational relaxation,
and mean square displacement, are anisotropic and depend on the distance from
the walls.
This research provides information needed for potential applications that use
molecular-scale motors in the complex confined geometries encountered in biology
and the laboratory.
\end{abstract}

\maketitle

\section{Introduction}
Synthetic self-propelled motors that convert chemical energy from their
environment into directed motion are examples of active objects with
distinctive and useful properties.
Much of the interest in such motors stems from potential applications that
exploit their directed motion to carry out functions involving cargo transport,
analogous to the active transport tasks performed by molecular motors in the
cell.%
~\cite{wang:13,jones:04,ozin:05,hong:10,kapral:13}

In many circumstances, motors operate in complex environments or systems confined
by boundaries.
Examples include motors used in microfluidic devices~\cite{garcia:13} and
potential {\it in vivo} applications such as targeted drug
delivery~\cite{patra:13,gao:14}.
Also, in many experiments currently being carried out, motors reside near a
surface as a result of gravitational or other forces, and the interactions of
motors with surfaces can lead to interesting dynamical effects.%
~\cite{palacci:10,bocquet:12,ozin:10,sengupta:14}
For these reasons, it is important to assess the influence of boundaries and
confinement on motor dynamics.

The effects of confinement on motor motion have been investigated previously.
The velocity of a spherical self-propelled particle with a point-like catalytic
site confined by a spherical wall has been computed analytically.%
~\cite{popescu:09}
Micron-scale Janus motors operating by self-diffusiophoresis have been studied
experimentally and numerically in circumstances where they collide with a wall,
are confined to a circular pore, and move through a triangular lattice of
obstacles forming a patterned environment.~\cite{volpe:11}
The dynamics of a self-diffusiophoretic Janus particle close to a hard wall has
been shown to exhibit reflection, steady sliding, and hovering.~\cite{uspal:14}
A study of such Janus particles near a wall has been carried out using continuum
hydrodynamics.~\cite{crowdy:13}
The rectification of a Brownian Janus particle in a triangular
channel~\cite{ghosh:13} and the escape of an ellipsoidal Brownian Janus
particle from a two-dimensional sinusoidally corrugated or square-shaped pore
have been simulated.~\cite{ghosh:14}
Studies of confined self-propelled rods have shown that they tend to reside
near surfaces.~\cite{elgeti:09}
The accumulation of self-propelled Brownian spheres near a hard wall has been
studied analytically and simulated using multi-particle collision
dynamics.~\cite{elgeti:13}
Similarly, ensembles of self-propelled colloidal rods aggregate and
structurally order on confining walls.~\cite{wensink:08}

In a related but somewhat different context, the effects of hydrodynamic
interactions on the confined motions of swimmers and biological organisms have
also been investigated.
Since biological organisms function in environments containing obstacles and
boundaries of various types, such studies are essential in order to obtain a
full understanding of swimming motions in a biological context.
Studies of hydrodynamic effects on self-propulsion of both synthetic swimmers
and biological organisms near boundaries and in complex media have been
carried out.%
~\cite{wensink:08,lauga:09,lauga:12,graham:05,graham:09,zoettl:14}
Boundaries can influence the collective behavior of swimmers and the collective
motions of swimmers confined to the gap between two planar surfaces have been
observed.%
~\cite{graham:05,graham:09}

Several factors must be considered in order to understand the effects of
confinement on chemically powered motors.
These include direct interactions with the boundaries, motor geometry, the
influence of boundaries on hydrodynamic flows, and the nature of chemical
gradients near walls.
At the micron to nanometer scales, self-propulsion is influenced by thermal
fluctuations.
Orientational Brownian motion limits the duration of the ballistic regime where
directed motion can be observed and leads to diffusive dynamics on long time
scales, albeit characterized by diffusion coefficients that reflect the
underlying active motion.
Interactions with boundaries can influence orientational Brownian motion and
modify motor behavior.

In this paper, we consider a regime of motor dynamics and confinement that
differs from that in earlier investigations.
We focus on very small {\AA}ngstr{\"o}m-scale motors confined in thin
nanometer-scale layers between two walls.
This study is prompted by several observations.
Recent experiments have demonstrated that catalytically active \SI{5}{\angstrom}
organometallic molecules~\cite{pavlick:13} as well as single enzyme
molecules~\cite{muddana:10,sengupta:13} display enhanced diffusion compared to
the same systems in the absence of chemical activity.
In addition, experimental studies of very small (\SI{30}{\nm}) synthetic
self-propelled Janus particles have also found enhanced diffusion
coefficients.~\cite{lee:14}
Although the precise mechanism responsible for diffusion enhancement is still a
matter of debate for the molecular systems, the observed phenomena are
correlated with chemical activity.
Such enzymatically induced motion has been reviewed recently.~\cite{gaspar:14}
In addition to these experimental results, molecular dynamics simulations of
chemically propelled {\AA}ngstr{\"o}m-size sphere-dimer motors that operate by
self-diffusiophoresis have shown that these motors can move distances of one to
several times their length before they reorient and have enhanced diffusion
coefficients.~\cite{colberg:14}

For such small-scale motors, whose size is comparable to the molecules
comprising the environment, solvent structural effects play an important role
and fluctuations dominate their behavior.
Nevertheless, the signatures of self-propulsion are clearly evident in
observable properties such as diffusion coefficients.
Enzymes carry out their functions in the complex cellular environment and,
potentially, very small synthetic motors will find applications on the cellular
level and other confined systems.
It is then important to consider effects due to the confinement on these very
small motors.

System-specific features of both the motor and solvent molecules can determine
the structural and dynamical properties of motors with small molecular-scale
dimensions.
Some general properties are common to all such motors.
Small motors reorient very rapidly, for example, nanometer-scale motors have
reorientation times of the order of nanoseconds or picoseconds, compared to
times of the order of seconds for micron-sized motors.
This time scale disparity will have an effect on the relative importance of
ballistic and diffusive motions.
For small motors, solvent structural correlations can influence effective
motor--wall forces in confined systems, as well as the concentrations of fuel
species in the motor vicinity.
Although the quantitative form these effects take depends on the system under
study, their presence is a common feature of molecular-scale motors.
In this study, we consider a simple dimer motor made from catalytic and
noncatalytic spheres in a dense atomic-like solvent in order to investigate
some of the general aspects of solvent and confinement effects on very small
motors.

The outline of the paper is as follows:
In Sec.~\ref{sec:model}, we describe the model for the motor and solvent
molecules, along with their interactions with the confining walls.
Section~\ref{sec:structure} presents results on static structural properties, in
particular, the nature of the motor position and its angular distributions
perpendicular and parallel to the walls.
Dynamical properties, including dimer velocity, orientational correlations, and
mean square displacement (MSD), are discussed in Sec.~\ref{sec:dynamics}.
The conclusions of the study are given in Sec.~\ref{sec:conc}.

\section{Confined self-propelled motors}\label{sec:model}

We consider motors that achieve propulsion by utilizing catalytic reactions
through a self-diffusiophoretic mechanism.%
~\cite{anderson:84,anderson:89,goles:05,kapral:13}
In this mechanism, a chemical reaction on a portion of the motor leads to
asymmetric distributions of reactants and products in the motor vicinity.
If the reactants and products interact with the motor through different
intermolecular potentials, this gives rise to a body force on the motor.
Since no external forces are applied to the system, the entire system, the motor
and its environment, are force-free, and momentum conservation leads to fluid
flows in the environment.
This mechanism is responsible for the directed self-propelled motion of such
motors.

The most widely studied motors of this type are Janus colloids where one face is
catalytic and the other chemically neutral.
Suppose that the chemical reaction $\text{A}\to\text{B}$ occurs on the catalytic
face and the A and B molecules interact with the Janus particle through
$U_{\text{J},\text{A}}$ and $U_{\text{J},\text{B}}$ intermolecular potentials.
Using the equations of continuum hydrodynamics, supplemented with boundary
conditions that account for reactions and fluid velocities on the surface of the
Janus particle, the velocity of the motor is given by
\begin{equation}
  V=(k_\text{B}T/\eta) \Lambda_\text{J} \langle \hat{\bm z}\cdot\nabla_\theta
  c_\text{B}(r_\text{J},\theta) \rangle_{{\mathcal S}},
\end{equation}
where $\hat{\bm z}$ is a unit vector from the noncatalytic to catalytic faces of
the Janus particle, $c_\text{B}(r_\text{J},\theta)$ is the concentration of
species B on the surface as a function of the angle $\theta$,
$\langle\cdots\rangle_{{\mathcal S}}$ denotes the average over the spherical
surface, and
\begin{equation}
  \Lambda_\text{J} = \int_0^\infty dr \;r[e^{-\beta U_{\text{J},\text{B}}(r)}
  -e^{-\beta U_{\text{J},\text{A}}(r)}],
\end{equation}
with $\beta=1/(k_\text{B}T)$.
From this expression, we see that the velocity depends on the surface average of
the gradient of the concentration along the surface, the difference in the
interaction potentials as determined by $\Lambda_\text{J}$ and $\eta$, the
viscosity of the solution.
These are features of all motors that operate by diffusiophoretic mechanisms.

Motor geometry is a factor to consider when dealing with wall interactions, and
instead of spherical Janus motors, we investigate the dynamics of sphere-dimer
motors that are simple examples of more complex, non-spherical motors.
A sphere-dimer consists of linked catalytic (C) and noncatalytic (N) spheres.
In this geometry, the catalytic region is restricted to one sphere.%
~\cite{kapral:07}
The directed motion of the motor is again due to a diffusiophoretic mechanism.
Experimental~\cite{ozin:10} and simulation~\cite{yuguo:08,kapral:14,yang:14}
studies of the motions of sphere-dimer motors in bulk solution have been
performed.
The analytic continuum theory for the velocity of a sphere-dimer motor is
considerably more involved due to the motor geometry, but it may be obtained
using a bispherical coordinate system.%
~\cite{popescu:11,reigh:15}
The theory predicts that the velocity of the motor along the dimer bond is given
by a form analogous to that for the Janus particle,
\begin{equation}
  V=(k_\text{B}T/\eta) \Lambda_{SD} K,
\end{equation}
with $K$ given by a complicated but explicit expression~\cite{reigh:15} that
plays the same role as the surface average of the concentration gradient for the
Janus particle.
If the intermolecular potentials only differ for interactions of A and B with
the noncatalytic sphere, $\Lambda_{SD}$ has the form
\begin{equation}
  \Lambda_{SD} = \int_0^\infty dr \;r[e^{-\beta U_{\text{N},\text{B}}(r)}
  -e^{-\beta U_{\text{N},\text{A}}(r)}].
\end{equation}

Since we are interested in very small {\AA}ngstr{\"o}m-scale motors, continuum
theory is not able to describe all relevant effects.
As noted earlier, fluctuations are especially important and the finite size of
the solvent molecules is comparable to that of the motor, both of which violate
the assumptions of a continuum description of the solvent.
Instead we adopt a full molecular dynamics description of a self-propelled
{\AA}ngstr{\"o}m-scale motor confined between two parallel walls.

In more detail, the sphere-dimer motor we consider consists of a catalytic C
sphere and a noncatalytic N~sphere, with diameters $\sigma_\text{C}=2\sigma$ and
$\sigma_\text{N}=4\sigma$, respectively, linked by a rigid bond of length $d$.
While various dimer-sphere sizes can be chosen, and the results will depend on
these values, we have selected a sphere-diameter ratio of
$\sigma_\text{C}/\sigma_\text{N} = 0.5$ that has been shown to yield the
highest propulsion velocity both experimentally~\cite{ozin:10} and
theoretically~\cite{reigh:15}.
The solvent consists of $N_\text{s}$ structureless A and B particles with
diameter $\sigma$ and mass $m$.
The solvent number density is $\varrho_\text{s} = N_\text{s}/V =
0.8\sigma^{-3}$, which corresponds to a dense fluid.
The sphere dimer masses, $M_\text{m} = \frac{\pi}{6} \varrho_\text{s} m
(\sigma_\text{C}^3 + \sigma_\text{N}^3)$, are chosen to make the dimer neutrally
buoyant.
All particles interact through a shifted, truncated Lennard-Jones potential,
$V_{ij}(r) = \epsilon_{ij}\bigl\{4\bigl[(\sigma_{ij}/r)^{12}
- (\sigma_{ij}/r)^6\bigr] + 1\bigr\}$
for $r < \sqrt[6]{2}\,\sigma_{ij}$ and zero otherwise.
Here $r$ is the distance between the centres of a pair of particles,
$\sigma_\text{CA} = \sigma_\text{CB} =
\frac{1}{2}\left(\sigma_\text{C}+\sigma\right)$ and $\sigma_\text{NA} =
\sigma_\text{NB} = \frac{1}{2}\left(\sigma_\text{N}+\sigma\right)$
for pairs of dimer sphere and solvent particle, and $\sigma_\text{AA} =
\sigma_\text{AB} = \sigma_\text{BB} = \sigma$ for pairs of solvent particles.
The interaction energy is $\epsilon$ for all pairs apart from NB pairs,
where $\epsilon_\text{NB} = 0.1\epsilon$ or $10\epsilon$.
The temperature of the system is $k_\text{B}T/\epsilon = 1$.

The system is contained in a slab with edge lengths $L_X = L_Y = 50\sigma$ and
$L_Z = 10,15,\dots,50\sigma$ with periodic boundary conditions in $X$ and $Y$
directions and two planar walls at $Z=0$ and $L_Z$.
The dimer spheres and solvent particles interact with the walls via a repulsive
9-3 Lennard-Jones potential, $V_S(\zeta)= \epsilon_{\text{w}S}\bigl\{(3\sqrt{3}/2)\,
\bigl[(\sigma_{\text{w}S}/\zeta)^9 - (\sigma_{\text{w}S}/\zeta)^3\bigr] + 1\bigr\}$
for $\zeta<\sqrt[6]{3}\,\sigma_{\text{w}S}$, where $\zeta$ is the distance to
the closest wall, and zero otherwise.
The wall interaction parameters are $\sigma_{\text{wA}} = \sigma_{\text{wB}} =
\sigma$, $\sigma_{\text{wC}} = (\sigma_\text{C}+\sigma)/2$, $\sigma_{\text{wN}} =
(\sigma_\text{N}+\sigma)/2$, $\epsilon_{\text{wA}} = \epsilon_{\text{wB}} = \epsilon$,
and $\epsilon_{\text{wC}} = \epsilon_{\text{wN}} = \epsilon$ or $2\epsilon$.

\begin{figure}[tbp]
  \begin{center}
    \includegraphics[width=\linewidth]{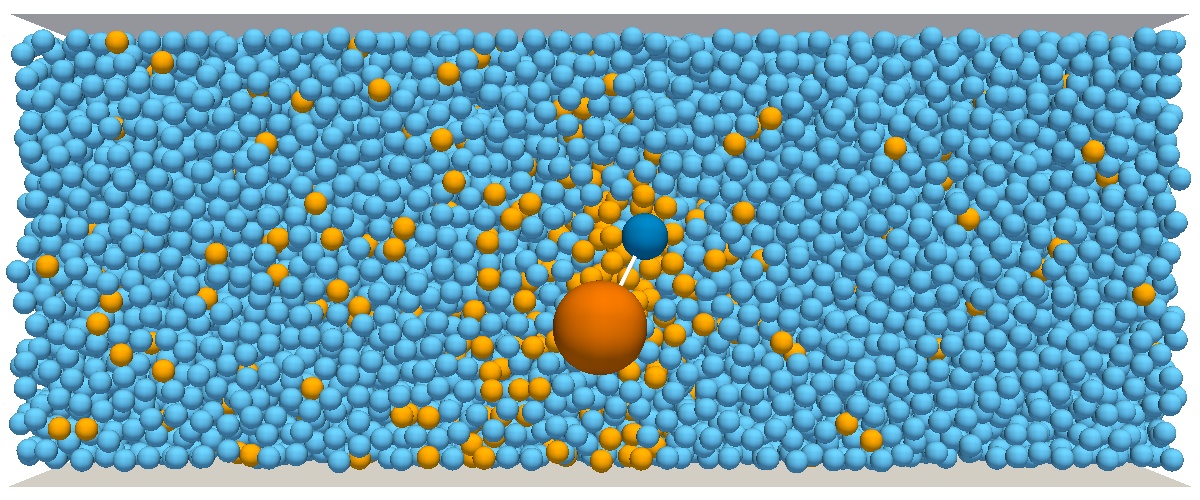}
  \end{center}
  \caption{Cross section of sphere-dimer motor with spheres C (blue) and N
  (vermilion) immersed in A (sky blue) and B (orange) solvent particles
  confined between two planar walls a distance $L_Z = 20$ apart.}
  \label{fig:dimer}
\end{figure}

The C~sphere catalyses the reaction A $\rightarrow$ B, where A is converted to B with
unit probability when it lies within a distance $\sqrt[6]{2}\,\sigma_\text{CB}$
from the C~sphere.
The system is maintained in a nonequilibrium steady state through a bulk
back-reaction B $\rightarrow$ A with rate $10^{-3}\,\tau^{-1}$ outside of the
C- and N-sphere interaction zones.
The dimer bond length is taken to be
$d = \sqrt[6]{2}\,(\sigma_\text{CB}+\sigma_\text{NB})$, which is chosen to
ensure energy conservation in the presence of reactions.

Full molecular dynamics~\footnote{The simulations were run on GPUs using a
massively parallel code written in OpenCL~C and Lua, which is distributed
under a free software license at \url{http://colberg.org/angstrom-dimer}.}
is used to follow the motions of the motor and solvent and capture their
structural and dynamical properties.~\footnote{In applications to experimental
systems the specific properties of the intermolecular potentials should be
taken into account.}
The model neglects the structure of the solvent molecules, but nevertheless
it is suited to reproduce structural and dynamical effects occurring at the
{\AA}ngstr{\"o}m scale.
As discussed in the Introduction, applications to experimental systems should
take into account the detailed structural properties of the motor and solvent
under investigation.
The velocity-Verlet integration time step for the molecular dynamics simulations
is $10^{-3}\tau$.
Simulation results are reported in dimensionless units with distance given in
units of $\sigma$, mass in units of $m$, energy in units of $\epsilon$, and time
in units of $\tau = \sigma\sqrt{m\epsilon^{-1}}$.

Parameters were chosen to model a dense fluid Argon-like solvent.
Given Argon~\cite{rahman:64} values of $\sigma = \SI{0.34}{\nm}$,
$\epsilon = \SI{120}{\K}\,k_\text{B}$, and $m = \SI{39.95}{\amu}$ and
$\tau = \SI{2.15}{\ps}$, we can assign physical values to our
{\AA}ngstr{\"o}m-scale motor simulations.
In these units the sphere-dimer monomers have radii of \SI{0.34}{\nm} and
\SI{0.68}{\nm} for the C and N~spheres, respectively.
The end-to-end length of the sphere-dimer motor is \SI{2.55}{\nm}.
The separation between the walls varies from \SI{3.4}{\nm} to \SI{17}{\nm}.

The continuum theory for sphere-dimer motors that operate by
self-diffusiophoresis has been carried out~\cite{popescu:11,reigh:15}
and the fluid flow fields associated with propulsion have been
determined~\cite{reigh:15}.
Motors with $\epsilon_\text{NB} = 0.1$ and small dimer bond lengths move in
the direction of the C~sphere and the far-field fluid flow is characteristic
of a puller.
For $\epsilon_\text{NB} = 10$, the motor is propelled in the direction of
the N~sphere and the far-field fluid flow is characteristic of a pusher.
Both types of motor were investigated in confinement between two planar walls,
and in bulk with periodic boundary conditions in all directions.

To isolate effects that are due to the self-propulsion of the motor, results
for an inactive dimer ($\epsilon_\text{NB} = 1$) are also presented.
Inactive dimers were constructed by turning off the chemical reactions that
power the motor and maintain the system in a nonequilibrium steady state.
The effect of the wall potential on the dimer is illustrated by varying the
repulsiveness between walls and spheres, denoted in the following as the
weaker ($\epsilon_{\text{w}S} = 1$) and the stronger wall potential
($\epsilon_{\text{w}S} = 2$).

An instantaneous configuration of the sphere dimer and surrounding A and B
solvent species confined between two walls is shown in Fig.~\ref{fig:dimer}.
In this figure one can see the relative sizes of dimer monomers and solvent
particles, as well as the inhomogeneous distribution of reactants and products
near the dimer and wall, which will play an important role in the subsequent
discussion.

\section{Structural ordering of dimer motor in confined geometry}\label{sec:structure}
\begin{figure}[tb]
  \begin{center}
    \input{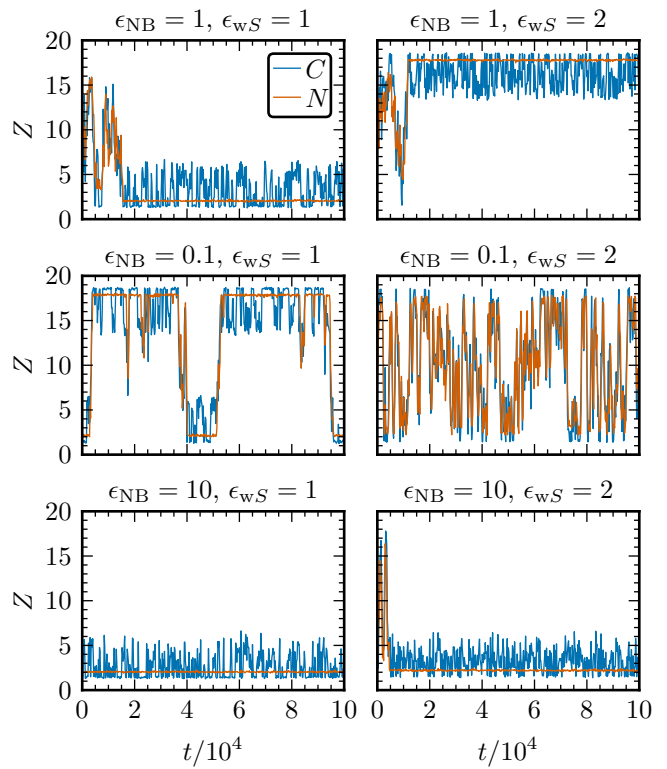}
  \end{center}
  \caption{Trajectories of the perpendicular distances of the N (red) and C
  (blue) dimer spheres to walls, $Z$, versus time for an inactive
  dimer (top), and puller (middle) and pusher motors (bottom).
  The left and right panels illustrate a weaker and a stronger repulsiveness
  of the wall-sphere potential, respectively.
  The separation between the walls is $L_Z = 20$.}
  \label{fig:position}
\end{figure}

We consider the structural ordering of the sphere-dimer motor when it is
confined between the two walls.
Sample trajectories of an inactive dimer, and puller and pusher motors
are shown in Fig.~\ref{fig:position}.
The dimer is initially placed along the middle plane between the walls.
Depending on the absence or presence of propulsion, the dimer takes a
long or short time, respectively, until its first contact with a wall.
In the case of the inactive dimer and the pusher motor, once the dimer has
reached a wall, it remains close to the wall for the remaining time.
The larger N~sphere resides at a fixed distance from the wall.
This effective attraction to the wall is explained by the strong solvent
depletion force~\cite{lekkerkerker:11} between the wall and the sphere.
The smaller C~sphere experiences a negligible depletion force, which allows
the sphere to freely explore the wall region that lies within the
constraints of the dimer bond.
The degree of repulsiveness of the wall has no discernible effect on the motion
of the inactive dimer or the pusher motor after being trapped by the wall.
The puller motor on the other hand behaves very differently depending on
the strength of the wall potential.
For the weaker wall repulsive potential, the puller motor is attached to the
wall most of the time as well.
In contrast to the inactive dimer and the pusher motor, however, the puller
motor occasionally detaches and makes transitions to the opposite wall or
returns to the original wall.
For the stronger wall repulsive potential, the puller motor is no longer
trapped by the walls and frequently makes transitions between them.

The different behaviors of the pusher versus puller motor are a
consequence of the respective direction of propulsion.
When the dimer bond is perpendicular to the wall, the propulsion force of
the pusher motor acts in the same direction as the depletion force on the
N~sphere, which reinforces the trapping by the wall.
The propulsion force of the puller motor opposes the depletion force,
which allows the motor to detach from the wall occasionally or frequently
depending on the wall repulsiveness.

\begin{figure}[tb]
  \begin{center}
    \input{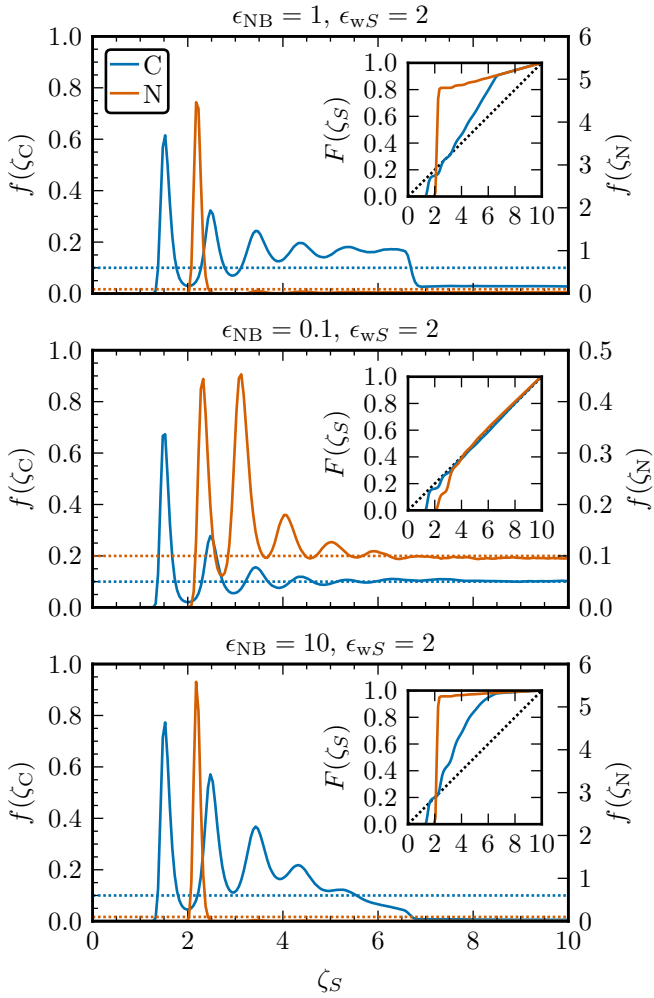}
  \end{center}
  \caption{Probability densities $f(\zeta_S)$ of the distance $\zeta_S$ between each
  dimer sphere $S$ and the nearest wall for an inactive dimer (top), and puller
  (middle) and pusher motors (bottom), for a wall separation $L_Z = 20$.
  The insets show the corresponding cumulative probability density $F(\zeta_S)$.
  The dotted lines indicate uniform density distributions.}
  \label{fig:density}
\end{figure}

The time- and ensemble-averaged~\footnote{Ensemble averages were carried out
over 30 realisations of each $10^8$ integration steps for a given system
$L_Z$, $\epsilon_\text{NB}$, and $\epsilon_{\text{w}S}$.} probability densities
of N and C~sphere positions are shown in Fig.~\ref{fig:density} for the
stronger wall potential for an inactive dimer, and puller and pusher motors.
The insets of Fig.~\ref{fig:density} show the probability densities
plotted as cumulative distributions.
The N~sphere density for the inactive dimer and the pusher motor exhibits
a pronounced peak, which is the signature of the trapping.
Since the time until first contact with a wall is shorter for the pusher
motor than for the inactive dimer, the peak magnitude of the N~sphere
density for the pusher motor is slightly larger and accounts for close to
100\% of the corresponding cumulative density.
The N~sphere density for the puller motor shows a richer structure with
damped oscillations that originate from the structural ordering of the
atomic-like solvent at the wall.
The C~sphere density has damped oscillations for all three cases.
Far from the wall, the C~sphere density converges to a uniform density
for the puller motor and decays to zero for the inactive dimer and the
pusher motor, identified earlier as the confinement by the dimer bond.

\begin{figure}[tb]
  \begin{center}
    \input{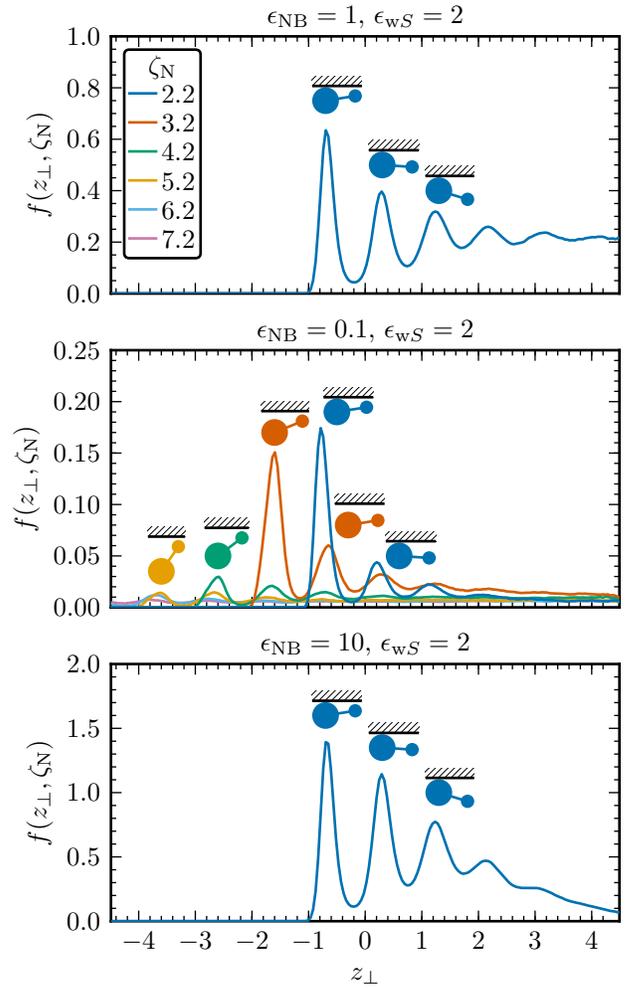}
  \end{center}
  \caption{The joint probability density $f(z_\perp, \zeta_\text{N})$ of the
  perpendicular component of the dimer bond vector, $z_\perp$, and the distance
  of the N~sphere from the closest wall, $\zeta_\text{N}$, for an inactive
  dimer (top), and puller (middle) and pusher motors (bottom), for a wall
  separation $L_Z = 20$.
  The color coding of the probability densities is indicated in the inset.
  The dimer configurations associated with the peak maxima are sketched in
  the figures.}
  \label{fig:angle}
\end{figure}

The structural ordering of the solvent also affects the average orientation of
the dimer relative to the walls.
We denote with ${\bf z}$ the bond vector pointing along the propulsion direction
from the centers of the N to C~spheres and separate the contributions
perpendicular and parallel to the wall, $z_\perp$ and ${\bf z}_\shortparallel$,
respectively.
The probability density of ${\bf z}_\shortparallel$ assumes a uniform distribution
due to the periodic boundary conditions in the $X$ and $Y$ directions.
The probability density of $z_\perp$ is strongly nonuniform, as seen in
Fig.~\ref{fig:angle}, which plots the joint probability density $f(z_\perp,
\zeta_\text{N})$ of the perpendicular component of the dimer bond vector,
$z_\perp$, and the distance of the N~sphere from the closest wall,
$\zeta_\text{N}$, for an inactive dimer, and puller and pusher motors.
These probability densities were computed by averaging over slabs
$\bigl[\zeta_N-\frac{\Delta\zeta}{2}, \zeta_N+\frac{\Delta\zeta}{2}\bigr]$ of
width $\Delta\zeta = 0.4$ parallel to the walls.

The orientation ranges from $z_\perp = -d$, which corresponds to the dimer
being perpendicular to the wall with the C~sphere oriented towards the wall,
to $z_\perp = d$, which corresponds to the opposite orientation.
The densities for all three cases show damped oscillations as a result of
solvent structural ordering close to the walls, with a period equal to the
diameter of a solvent molecule.
Each peak corresponds to a likely configuration of the dimer that is described
completely by the dimer bond orientation and the position of the N~sphere.
The joint probability density $f(z_\perp, \zeta_\text{N})$ plots allow one
to assign the motor configurations that contribute most significantly to
the peaks of the probability density.
These configurations are indicated in the figure.

For the inactive dimer and the pusher motor, the orientation densities resemble
the C~sphere position densities shown in Fig.~\ref{fig:density}, which, again,
reflects the trapping of the N~sphere by the wall.
For the inactive dimer, the density converges to a uniform distribution for
close-to-perpendicular orientations, while for the pusher motor, the density
decays to near zero.
This decay follows from the propulsion of the pusher motor parallel to the
wall, which causes the C~sphere to be pulled along by the trapped N~sphere
and thereby closer to the wall.
Compared to the inactive dimer and the pusher motor, the puller motor exhibits
a larger number of likely configurations, since the N~sphere is not trapped by
the wall.

The effects of confinement on suspensions of hydrodynamically interacting pusher
and puller swimmers were investigated by Hernandez-Ortiz et al.~\cite{graham:09}
The swimmers were modeled as dimers comprising beads linked by a stiff spring,
subject to a force that gives rise to propulsion.
Considering only very low dimer densities, the swimmer concentration profile as
a function of the distance from the walls has peaks near the walls.
As one moves farther from the walls, the concentration falls to close-to-zero
values for pushers and to non-zero values for pullers.
These findings are consistent with our puller and pusher single sphere-dimer
results.
However, we note that our simulations include a variety of other effects
including the self-generated chemical gradients, solvent structure and depletion
forces, fluctuations as well as hydrodynamics.

\section{Motor dynamics in confined geometry}\label{sec:dynamics}
\begin{figure}[tb]
  \begin{center}
    \input{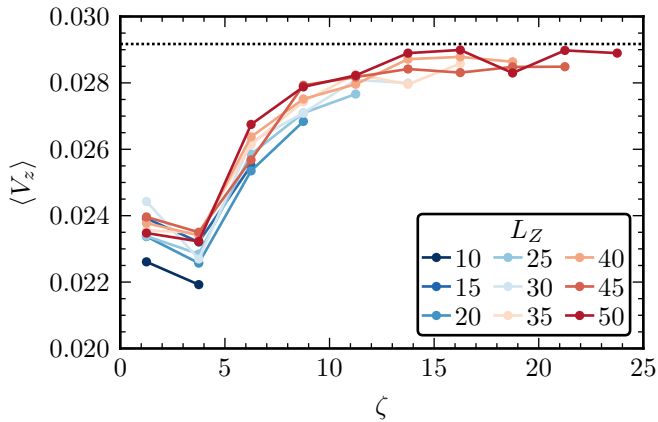}
  \end{center}
  \caption{Mean propulsion velocity, $\langle V_z\rangle$, of a confined
  puller motor as a function of the distance of its center of mass to the
  closest wall, $\zeta$, for $\epsilon_{\text{w}S}$ = 2.
  The velocities are averaged over slabs
  $\left[\zeta-\frac{\Delta\zeta}{2}, \zeta+\frac{\Delta\zeta}{2}\right]$
  of width $\Delta\zeta = 2.5$ parallel to the walls.
  The dotted line shows the propulsion velocity of an unconfined puller motor.}
  \label{fig:velocity}
\end{figure}
The catalytic reaction at the C~sphere is responsible for a propulsion force
along the dimer bond, which, for our choice of interaction parameters, results in
a mean motor velocity, $\langle V_z\rangle\neq0$.
The propulsion velocity of a puller motor is shown in Fig.~\ref{fig:velocity}
as a function of the distance from the wall for various wall separations, $L_Z$.
Close to the wall dimer propulsion is suppressed leading to smaller mean
velocities of $\langle V_z\rangle\approx 0.023$.
Far from the wall the velocity converges to a limit value $\langle
V_z\rangle\approx 0.029$, which corresponds to the propulsion velocity
of an unconfined puller motor.
A dimer confined in a narrower slab with $L_Z\leq 20$ has a slightly
smaller velocity close to the wall than a dimer confined in a wider slab, but
overall the velocity profile is roughly independent of $L_Z$.

The dimer is subject to strong thermal fluctuations from the solvent that affect
its translational and rotational motion.
In particular, rotational Brownian motion will reorient the dimer motor so that
its ballistic motion will only be manifested for times less than the average
reorientation time.
(In fact, for these {\AA}ngstr{\"o}m-size dimer motors, the ballistic regime
is dominated by thermal inertial effects.~\cite{colberg:14})
For times longer than this, the behavior will be diffusive but with an enhanced
diffusion coefficient.
As discussed earlier, these reorientation effects are especially strong for
{\AA}ngstr{\"o}m-size dimers and the crossover from ballistic to diffusive
motion occurs on the time scale of picoseconds.
For a dimer motor confined by two parallel walls, rotational motion is no longer
isotropic and it is interesting to examine how confinement influences this
property.

\begin{figure}[tb]
  \begin{center}
    \input{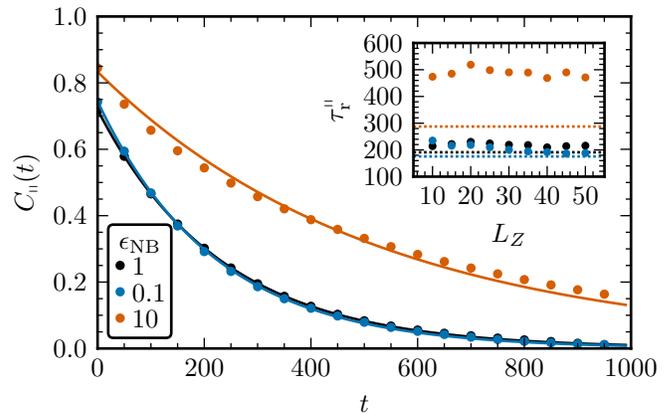}
  \end{center}
  \caption{Parallel orientational correlation function of a confined
    inactive dimer, and confined puller and pusher motors, for $L_Z = 20$
    and $\epsilon_{\text{w}S} = 2$.
    The solid lines show fits to an exponential decay.
    The inset shows the fitted relaxation times $\tau_\text{r}^\shortparallel$
    as a function of $L_Z$.
    The dotted lines indicate $\tau_\text{r}$ for the respective bulk case.}
  \label{fig:orientation}
\end{figure}
If, as earlier, we denote by $\hat{\bf z}$ the unit vector along the dimer bond,
we may resolve this quantity into its components perpendicular and parallel to
the wall, $\hat{z}_\perp$ and $\hat{\bf z}_\shortparallel$, respectively.
Here we focus on the parallel component since the corresponding correlation
function may be compared directly with that of an unconfined dimer.
The parallel orientational correlation function,
$C_\shortparallel(t) = \langle \hat{\bf z}_\shortparallel(t) \cdot
\hat{\bf z}_\shortparallel \rangle$, is plotted in Fig.~\ref{fig:orientation}
versus time for an inactive dimer, and puller and pusher motors.
The parallel orientational correlation follows an exponential decay,
$C_\shortparallel(t)\propto{\mathrm e}^{-t/\tau_\text{r}^\shortparallel}$, and the
fitted relaxation times $\tau_\text{r}^\shortparallel$ are shown in the inset of
Fig.~\ref{fig:orientation} as a function of the wall separation $L_Z$.
For the inactive dimer and the puller motor, the confined and unconfined
dimers exhibit roughly the same reorientation time.
For the pusher motor, however, the relaxation times of the confined dimer
are almost twice as large as for the unconfined dimer.
The confinement between the two walls has a negligible effect on the
parallel orientational correlation in the case of the inactive dimer and
the puller motor, but a strong effect in the case of the pusher motor.
We see this effect for both values of $\epsilon_{\text{w}S}$.

As discussed above, one of the major, easily obtainable signatures of
propulsion for very small motors is the existence of enhanced diffusion.
Confinement will influence this transport property.
We have shown earlier~\cite{colberg:14} that the MSD
of an unconfined {\AA}ngstr{\"o}m-scale sphere-dimer motor exhibits an enhanced
diffusive regime and is approximated well by the equation ($d$ is now the number
of dimensions)
\begin{eqnarray}
  \label{eq:MSD-th}
  \Delta L^2\left(t\right)
   &=& 2dD_\text{m}t
  - 2\langle V_z\rangle^2\tau_\text{r}^2
  \left(1-{\mathrm e}^{-t/\tau_\text{r}}\right)
  \\
  &&\qquad\qquad
  - 2d\frac{k_\text{B}T}{M_\text{m}}\tau_\text{v}^2
  \left(1-{\mathrm e}^{-t/\tau_\text{v}}\right)\nonumber.
\end{eqnarray}
Here $\tau_\text{v}$ is the decay time of the velocity fluctuations
as determined from the diffusion coefficient of an inactive dimer,
$D_0 = (k_\text{B}T/M_\text{m})\,\tau_\text{v}$.
The MSD reduces to $\Delta L^2(t) \approx (dk_\text{B}T/M_\text{m}
+ \langle V_z\rangle^2)\,t^2$ in the ballistic regime, $t\ll\tau_\text{v}$,
and to $\Delta L^2(t) \approx 2d\,(D_0 +
\frac{1}{d}\langle V_z\rangle^2\tau_\text{r})\,t = 2dD_\text{m}t$ in the
diffusive regime, $t\gg\tau_\text{r}$.~\cite{colberg:14}

\begin{figure}[tb]
  \begin{center}
    \input{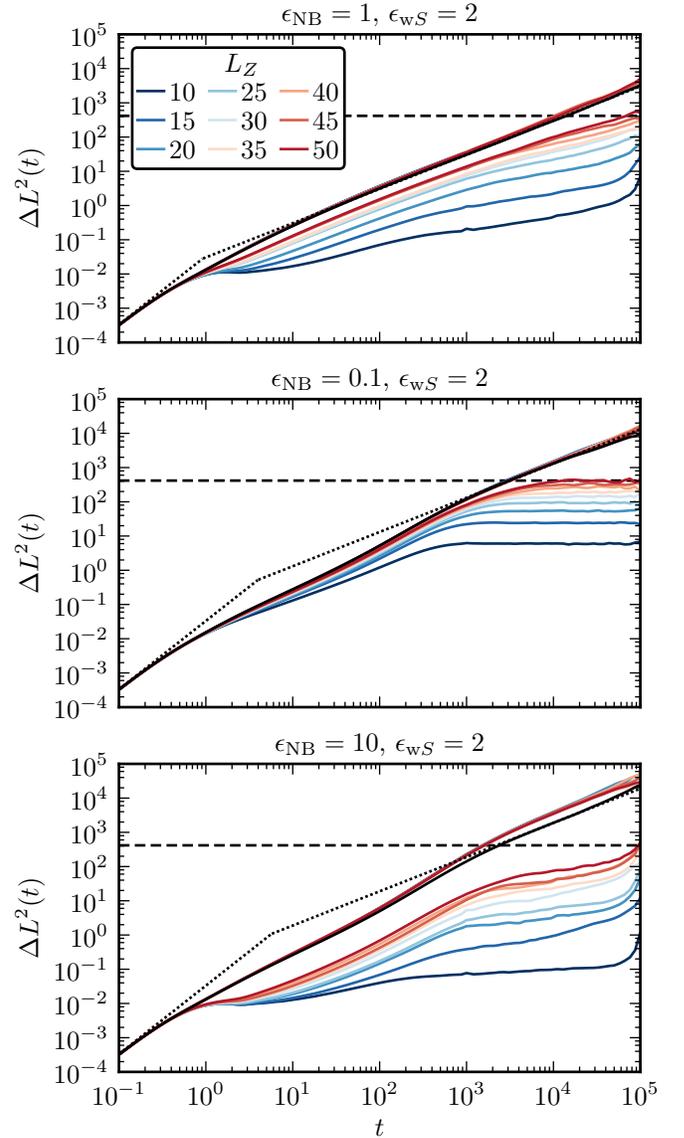}
  \end{center}
  \caption{One-dimensional mean square displacements parallel and perpendicular
  to walls depending on wall separation $L_Z$ for an inactive dimer (top),
  and puller (middle) and pusher motors (bottom).
  In each panel, the ensemble of overlapping curves at the top corresponds to
  the parallel MSDs, and the diverging curves below correspond to the
  perpendicular MSDs.
  The solid black line corresponds to the one-dimensional MSD for an
  unconfined dimer.
  The dotted lines show the ballistic and diffusive regimes of the
  one-dimensional MSD for an unconfined dimer.
  The dashed, horizontal line indicates the theoretical limit of confined
  diffusion for $L_Z = 50$.}
  \label{fig:msd}
\end{figure}

The MSDs of a confined inactive dimer, and confined puller and pusher
motors are shown in Fig.~\ref{fig:msd}.
\begin{table}[tb]
  \centering
  \begin{ruledtabular}
  \begin{tabular}{cccrrrr}
    $L_Z$
    & $\epsilon_\text{NB}$
    & d
    & \multicolumn{1}{c}{$\langle V_z\rangle$}
    & \multicolumn{1}{c}{$\tau_\text{r}^\shortparallel$}
    & \multicolumn{1}{c}{$D_\text{m}^\shortparallel$}
    & \multicolumn{1}{c}{$\Delta D_{\text{m},\text{th}}^\shortparallel$}
    \\
    \hline
    10 & 1 & & & 214 & 0.019 & \\
    10 & 0.1 & 2 & 0.022 & 235 & 0.075 & +1\% \\
    10 & 10 & 2 & -0.022 & 474 & 0.192 & -31\% \\
    \hline
    20 & 1 & & & 230 & 0.017 & \\
    20 & 0.1 & 3 & 0.024 & 219 & 0.069 & -12\% \\
    20 & 10 & 2 & -0.021 & 519 & 0.199 & -34\% \\
    \hline
    30 & 1 & & & 219 & 0.018 & \\
    30 & 0.1 & 3 & 0.026 & 201 & 0.066 & -5\% \\
    30 & 10 & 2 & -0.021 & 490 & 0.180 & -28\% \\
    \hline
    40 & 1 & & & 209 & 0.017 & \\
    40 & 0.1 & 3 & 0.027 & 192 & 0.074 & -14\% \\
    40 & 10 & 2 & -0.022 & 469 & 0.184 & -31\% \\
    \hline
    50 & 1 & & & 215 & 0.018 & \\
    50 & 0.1 & 3 & 0.027 & 188 & 0.070 & -8\% \\
    50 & 10 & 2 & -0.022 & 471 & 0.175 & -26\% \\
    \hline
    $\infty$ & 1 & & & 191 & 0.015 & \\
    $\infty$ & 0.1 & 3 & 0.029 & 176 & 0.066 & -1\% \\
    $\infty$ & 10 & 3 & -0.025 & 287 & 0.095 & -23\% \\
  \end{tabular}
  \end{ruledtabular}
  \caption{Dynamical properties for an inactive dimer, and puller and
    pusher motors, for various wall separations $L_Z$ and
    $\epsilon_{\text{w}S} = 2$, and the bulk case:
    Mean propulsion velocity, $\langle V_z\rangle$, relaxation time of the
    parallel orientational correlation, $\tau_\text{r}^\shortparallel$,
    diffusion constant of the parallel mean-square displacement,
    $D_\text{m}^\shortparallel$, and deviation of the theoretical estimate
    for the diffusion constant, $D_{\text{m},\text{th}}^\shortparallel =
    D_0 + \frac{1}{d}\langle V_z\rangle^2\tau_\text{r}^\shortparallel$,
    where $d$ is the effective number of dimensions of the system that
    minimizes $\Delta D_{\text{m},\text{th}}^\shortparallel$.}
  \label{tab:dimer}
\end{table}
To account for the confined geometry, the MSDs are separated into components
perpendicular and parallel to the walls, $\Delta L_\perp^2(t) =
\langle(R_\perp(t)-R_\perp)^2\rangle$ and $\frac{1}{2}\Delta L_\shortparallel^2(t) =
\frac{1}{2}\langle |{\bf R}_\shortparallel(t) - {\bf R}_\shortparallel|^2\rangle$,
where ${\bf R}(t)$ is the center of mass of the dimer.
For the inactive dimer and the puller motor, the parallel components show
negligible dependence on $L_Z$ and overlap with the one-dimensional MSD of
the respective bulk case.
For the pusher motor, the parallel components also show negligible dependence
on $L_Z$ but deviate significantly from the MSD of the bulk case.
In contrast to the parallel components, the perpendicular components show
$L_Z$-dependent behavior.
For the inactive dimer, the perpendicular MSDs reach a diffusive regime for
intermediate times for all but the smallest $L_Z = 10$, and the diffusion
coefficient decreases with decreasing $L_Z$.
For the puller motor, the perpendicular MSDs show the same qualitative
behavior as the bulk MSD in the intermediate time regime; in the long-time
limit, however, where the bulk MSD reaches the enhanced diffusive regime,
the confined MSDs instead converge to $L_Z$-dependent limits, which are in good
agreement with the theoretical limit $\frac{1}{6}\Delta L_Z^2$ of confined
diffusion for a one-dimensional random walk.~\cite{bickel:07}
For the pusher motor, the perpendicular MSDs show superdiffusive behavior
in the intermediate time regime; compared to the puller motor, however,
the superdiffusive regime is strongly suppressed, and in the long-time limit,
the MSDs become subdiffusive, to the point of a plateau for the smallest
$L_Z = 10$ that is two orders of magnitude below $\frac{1}{6}\Delta L_Z^2$.
The differences in the $L_Z$-dependence of the intermediate regimes of the
perpendicular MSDs for the three cases follow from the different structural
ordering of the dimer.
As noted before, the puller motor frequently makes transitions between the walls
for $\epsilon_{\text{w}S} = 2$, which results in long periods away from the
walls with superdiffusive motion, and short periods close to the walls with
suppressed motion perpendicular to the walls.
The inactive dimer and the pusher motor, on the other hand, are trapped
for very long periods (as long as our simulation times) after their first
contact with a wall; their perpendicular MSDs reflect the averaging over the
initial period of diffusive or superdiffusive motion, respectively, and the
subsequent period of suppressed motion perpendicular to the walls.
The time until first contact grows with $L_Z$ and is larger for the inactive
dimer than the motor dimers due to the absence versus presence of propulsion,
which explains the more strongly suppressed perpendicular MSDs for the pusher
motor compared to the inactive dimer.

The diffusion constants extracted from the parallel MSDs of confined dimers and
the MSD of unconfined dimers are listed in Table~\ref{tab:dimer}; along with
the motor velocities, the reorientation times parallel to the walls, and
theoretical estimates for the diffusion constant from these quantities.
For the inactive dimer and the puller motor, the diffusion constants
$D_\text{m}^\shortparallel$ of the confined and unconfined dimers are approximately
equal; for the pusher motor, the diffusion constant of the confined dimer is
twice that of the unconfined dimer.
The effective number of dimensions $d$ in the theoretical estimate for the
diffusion constant, $D_\text{m}^\shortparallel =
D_0 + \frac{1}{d}\langle V_z\rangle^2\tau_\text{r}^\shortparallel$,
is chosen as 2 for the pusher motor, which is trapped by a wall most of the
time, and 2 or 3, depending on $L_Z$, for the puller motor, which for small
$L_Z$ spends most of the time in the wall regions.
For the puller motor, the theoretical values of $D_\text{m}^\shortparallel$ are in
good agreement with the measured values.
For the pusher motor, the theoretical values consistently underestimate the
measured values, which, however, is also the case for the unconfined pusher
motor; hence, this deviation is not specific to the confinement.
Since the reorientation times for the pusher motor deviate substantially from
those of the inactive dimer, this indicates that reorientation also has an
active component.
The simple estimate for diffusion enhancement assumes simple inactive
reorientation and this is likely the origin of the quantitative failure of the
simple theoretical estimate for the pusher motor.
Nevertheless, these considerations suggest that the increased
$\tau_\text{r}^\shortparallel$ of the confined pusher motor is responsible for the
increased $D_\text{m}^\shortparallel$ compared to the bulk case.

\begin{figure}[tb]
  \begin{center}
    \input{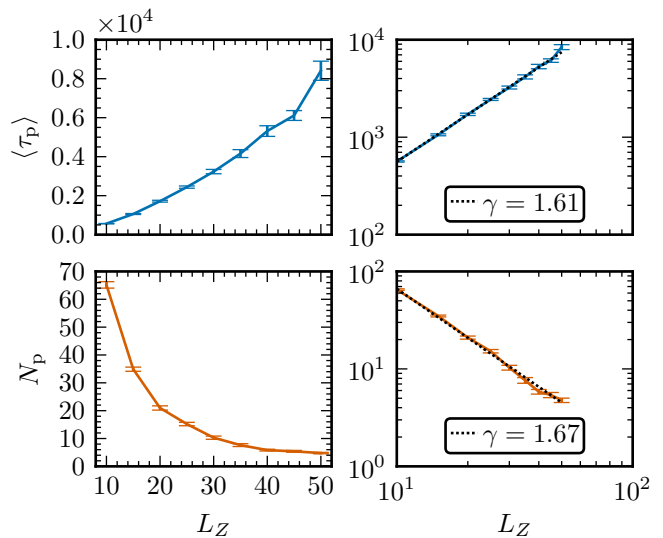}
  \end{center}
  \caption{Mean first-passage time (top) and number of passages (bottom)
  of a puller motor between opposite wall regions as a function of $L_Z$,
  for $\epsilon_{\text{w}S}$ = 2.
  The center of mass of the dimer is considered to be in a wall region
  when the distance from the wall $\zeta < 2.5$.
  The right panels show fits to power laws, $\langle\tau_\text{p}\rangle
  \propto L_Z^\gamma$ and $N_\text{p}\propto L_Z^{-\gamma}$.}
  \label{fig:passage}
\end{figure}

For the puller motor that makes frequent transitions between the walls, the mean
first-passage time and the number of passages are plotted as a function of
$L_Z$ in Fig.~\ref{fig:passage}.
The first-passage time and the number of passages are found to follow power
laws, $\langle\tau_\text{p}\rangle \propto L_Z^\gamma$ and $N_\text{p}\propto
L_Z^{-\gamma}$, respectively, with exponent $\gamma\approx 1.6$, which indicates
superdiffusive transitions between the walls.~\footnote{A comparable scaling
behavior has been derived for an unrelated system with superdiffusive
transport, where the first-passage time is found to follow a power law with
exponent $\frac{4}{3}$.~\cite{redner:90}}

\section{Conclusion}\label{sec:conc}
Molecular motors effect active transport in the cell and carry out a variety of
other cellular functions.
It is certainly interesting to imagine that, in the future, synthetic
molecular-scale motors and machines could also be used for biological functions
at the cellular level.
The fact that the dynamical properties of small catalysts and single enzyme
molecules are modified by chemical activity supports such a notion.%
~\cite{pavlick:13,muddana:10,sengupta:13}
While the potential applications of micron-scale motors are being actively
explored and many of their properties can be modeled through continuum theories,
the dynamics of {\AA}ngstr{\"o}m-size synthetic motors involve some additional
considerations that can only be captured through full molecular dynamics
simulations.
They are dominated by fluctuations, molecular reorientation often occurs on
picosecond time scales limiting the regime where ballistic motion dominates,
solvent sizes are often comparable to those of the motor, so structural effects
come into play and solvent depletion forces are very strong.
These features, combined with the fact that the utility of a continuum
description of the solvent velocity flow fields must be confirmed, make these
active objects challenging to study.

On the {\AA}ngstr{\"o}m scale, specific characteristics of the molecular nature
of the motor and its environment become important.
This implies that studies of real systems on these microscopic scales will have
to account for these molecular details.
The model system studied in this paper provides insight into some of the
qualitative characteristics that such systems may exhibit.
In particular, the results show how the distinctive features of
{\AA}ngstr{\"o}m-scale motors, in combination with confinement, change motor
spatial structure and dynamics.
We have seen that self-propulsion is able to counteract solvent depletion forces
in some circumstances so that the structural ordering of inactive dimers
between the walls is very different from that of active dimer motors.
The strong anisotropy and spatial dependence of dynamical properties including
motor velocity, orientation, and diffusion have their origin in the confining
geometry.
Such information should prove useful for applications involving very small
motors in cellular or microfluidic environments.

\begin{acknowledgments}
This work was supported in part by a grant from the Natural Sciences and
Engineering Research Council of Canada and Compute Canada.
\end{acknowledgments}

\bibliographystyle{apsrev4-1}
\bibliography{confined_dimer}

\begin{thebibliography}{49}%
\makeatletter
\providecommand \@ifxundefined [1]{%
 \@ifx{#1\undefined}
}%
\providecommand \@ifnum [1]{%
 \ifnum #1\expandafter \@firstoftwo
 \else \expandafter \@secondoftwo
 \fi
}%
\providecommand \@ifx [1]{%
 \ifx #1\expandafter \@firstoftwo
 \else \expandafter \@secondoftwo
 \fi
}%
\providecommand \natexlab [1]{#1}%
\providecommand \enquote  [1]{``#1''}%
\providecommand \bibnamefont  [1]{#1}%
\providecommand \bibfnamefont [1]{#1}%
\providecommand \citenamefont [1]{#1}%
\providecommand \href@noop [0]{\@secondoftwo}%
\providecommand \href [0]{\begingroup \@sanitize@url \@href}%
\providecommand \@href[1]{\@@startlink{#1}\@@href}%
\providecommand \@@href[1]{\endgroup#1\@@endlink}%
\providecommand \@sanitize@url [0]{\catcode `\\12\catcode `\$12\catcode
  `\&12\catcode `\#12\catcode `\^12\catcode `\_12\catcode `\%12\relax}%
\providecommand \@@startlink[1]{}%
\providecommand \@@endlink[0]{}%
\providecommand \url  [0]{\begingroup\@sanitize@url \@url }%
\providecommand \@url [1]{\endgroup\@href {#1}{\urlprefix }}%
\providecommand \urlprefix  [0]{URL }%
\providecommand \Eprint [0]{\href }%
\providecommand \doibase [0]{http://dx.doi.org/}%
\providecommand \selectlanguage [0]{\@gobble}%
\providecommand \bibinfo  [0]{\@secondoftwo}%
\providecommand \bibfield  [0]{\@secondoftwo}%
\providecommand \translation [1]{[#1]}%
\providecommand \BibitemOpen [0]{}%
\providecommand \bibitemStop [0]{}%
\providecommand \bibitemNoStop [0]{.\EOS\space}%
\providecommand \EOS [0]{\spacefactor3000\relax}%
\providecommand \BibitemShut  [1]{\csname bibitem#1\endcsname}%
\let\auto@bib@innerbib\@empty
\bibitem [{\citenamefont {Wang}(2013)}]{wang:13}%
  \BibitemOpen
  \bibfield  {author} {\bibinfo {author} {\bibfnamefont {J.}~\bibnamefont
  {Wang}},\ }\href@noop {} {\emph {\bibinfo {title} {Nanomachines: Fundamentals
  and Applications}}}\ (\bibinfo  {publisher} {Wiley-VCH},\ \bibinfo {address}
  {Weinheim},\ \bibinfo {year} {2013})\BibitemShut {NoStop}%
\bibitem [{\citenamefont {Jones}(2004)}]{jones:04}%
  \BibitemOpen
  \bibfield  {author} {\bibinfo {author} {\bibfnamefont {R.~A.~L.}\
  \bibnamefont {Jones}},\ }\href@noop {} {\emph {\bibinfo {title} {Soft
  Machines: Nanotechnology and Life}}}\ (\bibinfo  {publisher} {Oxford
  University Press},\ \bibinfo {address} {Oxford},\ \bibinfo {year}
  {2004})\BibitemShut {NoStop}%
\bibitem [{\citenamefont {Ozin}\ \emph {et~al.}(2005)\citenamefont {Ozin},
  \citenamefont {Manners}, \citenamefont {Fournier-Bidoz},\ and\ \citenamefont
  {Arsenault}}]{ozin:05}%
  \BibitemOpen
  \bibfield  {author} {\bibinfo {author} {\bibfnamefont {G.~A.}\ \bibnamefont
  {Ozin}}, \bibinfo {author} {\bibfnamefont {I.}~\bibnamefont {Manners}},
  \bibinfo {author} {\bibfnamefont {S.}~\bibnamefont {Fournier-Bidoz}}, \ and\
  \bibinfo {author} {\bibfnamefont {A.}~\bibnamefont {Arsenault}},\ }\href
  {\doibase 10.1002/adma.200501767} {\bibfield  {journal} {\bibinfo  {journal}
  {Adv. Mater.}\ }\textbf {\bibinfo {volume} {17}},\ \bibinfo {pages} {3011}
  (\bibinfo {year} {2005})}\BibitemShut {NoStop}%
\bibitem [{\citenamefont {Hong}\ \emph {et~al.}(2010)\citenamefont {Hong},
  \citenamefont {Velegol}, \citenamefont {Chaturvedi},\ and\ \citenamefont
  {Sen}}]{hong:10}%
  \BibitemOpen
  \bibfield  {author} {\bibinfo {author} {\bibfnamefont {Y.}~\bibnamefont
  {Hong}}, \bibinfo {author} {\bibfnamefont {D.}~\bibnamefont {Velegol}},
  \bibinfo {author} {\bibfnamefont {N.}~\bibnamefont {Chaturvedi}}, \ and\
  \bibinfo {author} {\bibfnamefont {A.}~\bibnamefont {Sen}},\ }\href {\doibase
  10.1039/b917741h} {\bibfield  {journal} {\bibinfo  {journal} {Phys. Chem.
  Chem. Phys.}\ }\textbf {\bibinfo {volume} {12}},\ \bibinfo {pages} {1423}
  (\bibinfo {year} {2010})}\BibitemShut {NoStop}%
\bibitem [{\citenamefont {Kapral}(2013)}]{kapral:13}%
  \BibitemOpen
  \bibfield  {author} {\bibinfo {author} {\bibfnamefont {R.}~\bibnamefont
  {Kapral}},\ }\href {\doibase 10.1063/1.4773981} {\bibfield  {journal}
  {\bibinfo  {journal} {J. Chem. Phys.}\ }\textbf {\bibinfo {volume} {138}},\
  \bibinfo {pages} {020901} (\bibinfo {year} {2013})}\BibitemShut {NoStop}%
\bibitem [{\citenamefont {Garc{\'i}a}\ \emph {et~al.}(2013)\citenamefont
  {Garc{\'i}a}, \citenamefont {Orozco}, \citenamefont {Guix}, \citenamefont
  {Gao}, \citenamefont {Sattayasamitsathit}, \citenamefont {Escarpa},
  \citenamefont {Merko{\c c}i},\ and\ \citenamefont {Wang}}]{garcia:13}%
  \BibitemOpen
  \bibfield  {author} {\bibinfo {author} {\bibfnamefont {M.}~\bibnamefont
  {Garc{\'i}a}}, \bibinfo {author} {\bibfnamefont {J.}~\bibnamefont {Orozco}},
  \bibinfo {author} {\bibfnamefont {M.}~\bibnamefont {Guix}}, \bibinfo {author}
  {\bibfnamefont {W.}~\bibnamefont {Gao}}, \bibinfo {author} {\bibfnamefont
  {S.}~\bibnamefont {Sattayasamitsathit}}, \bibinfo {author} {\bibfnamefont
  {A.}~\bibnamefont {Escarpa}}, \bibinfo {author} {\bibfnamefont
  {A.}~\bibnamefont {Merko{\c c}i}}, \ and\ \bibinfo {author} {\bibfnamefont
  {J.}~\bibnamefont {Wang}},\ }\href {\doibase 10.1039/C2NR32400H} {\bibfield
  {journal} {\bibinfo  {journal} {Nanoscale}\ }\textbf {\bibinfo {volume}
  {5}},\ \bibinfo {pages} {1325} (\bibinfo {year} {2013})}\BibitemShut
  {NoStop}%
\bibitem [{\citenamefont {D.~Patra}\ \emph {et~al.}(2013)\citenamefont
  {D.~Patra}, \citenamefont {Duan}, \citenamefont {Zhang}, \citenamefont
  {Pavlick},\ and\ \citenamefont {Sen}}]{patra:13}%
  \BibitemOpen
  \bibfield  {author} {\bibinfo {author} {\bibfnamefont {S.~S.}\ \bibnamefont
  {D.~Patra}}, \bibinfo {author} {\bibfnamefont {W.}~\bibnamefont {Duan}},
  \bibinfo {author} {\bibfnamefont {H.}~\bibnamefont {Zhang}}, \bibinfo
  {author} {\bibfnamefont {R.}~\bibnamefont {Pavlick}}, \ and\ \bibinfo
  {author} {\bibfnamefont {A.}~\bibnamefont {Sen}},\ }\href {\doibase
  10.1039/C2NR32600K} {\bibfield  {journal} {\bibinfo  {journal} {Nanoscale}\
  }\textbf {\bibinfo {volume} {5}},\ \bibinfo {pages} {1273} (\bibinfo {year}
  {2013})}\BibitemShut {NoStop}%
\bibitem [{\citenamefont {Gao}\ and\ \citenamefont {Wang}(2014)}]{gao:14}%
  \BibitemOpen
  \bibfield  {author} {\bibinfo {author} {\bibfnamefont {W.}~\bibnamefont
  {Gao}}\ and\ \bibinfo {author} {\bibfnamefont {J.}~\bibnamefont {Wang}},\
  }\href {\doibase 10.1039/C4NR03124E} {\bibfield  {journal} {\bibinfo
  {journal} {Nanoscale}\ }\textbf {\bibinfo {volume} {6}},\ \bibinfo {pages}
  {10486} (\bibinfo {year} {2014})}\BibitemShut {NoStop}%
\bibitem [{\citenamefont {Palacci}\ \emph {et~al.}(2010)\citenamefont
  {Palacci}, \citenamefont {Cottin-Bizonne}, \citenamefont {Ybert},\ and\
  \citenamefont {Bocquet}}]{palacci:10}%
  \BibitemOpen
  \bibfield  {author} {\bibinfo {author} {\bibfnamefont {J.}~\bibnamefont
  {Palacci}}, \bibinfo {author} {\bibfnamefont {C.}~\bibnamefont
  {Cottin-Bizonne}}, \bibinfo {author} {\bibfnamefont {C.}~\bibnamefont
  {Ybert}}, \ and\ \bibinfo {author} {\bibfnamefont {L.}~\bibnamefont
  {Bocquet}},\ }\href {\doibase 10.1103/PhysRevLett.105.088304} {\bibfield
  {journal} {\bibinfo  {journal} {Phys. Rev. Lett.}\ }\textbf {\bibinfo
  {volume} {105}},\ \bibinfo {pages} {088304} (\bibinfo {year}
  {2010})}\BibitemShut {NoStop}%
\bibitem [{\citenamefont {Theurkauff}\ \emph {et~al.}(2012)\citenamefont
  {Theurkauff}, \citenamefont {Cottin-Bizonne}, \citenamefont {Palacci},
  \citenamefont {Ybert},\ and\ \citenamefont {Bocquet}}]{bocquet:12}%
  \BibitemOpen
  \bibfield  {author} {\bibinfo {author} {\bibfnamefont {I.}~\bibnamefont
  {Theurkauff}}, \bibinfo {author} {\bibfnamefont {C.}~\bibnamefont
  {Cottin-Bizonne}}, \bibinfo {author} {\bibfnamefont {J.}~\bibnamefont
  {Palacci}}, \bibinfo {author} {\bibfnamefont {C.}~\bibnamefont {Ybert}}, \
  and\ \bibinfo {author} {\bibfnamefont {L.}~\bibnamefont {Bocquet}},\ }\href
  {\doibase 10.1103/PhysRevLett.108.268303} {\bibfield  {journal} {\bibinfo
  {journal} {Phys. Rev. Lett.}\ }\textbf {\bibinfo {volume} {108}},\ \bibinfo
  {pages} {268303} (\bibinfo {year} {2012})}\BibitemShut {NoStop}%
\bibitem [{\citenamefont {Valadares}\ \emph {et~al.}(2010)\citenamefont
  {Valadares}, \citenamefont {Tao}, \citenamefont {Zacharia}, \citenamefont
  {Kitaev}, \citenamefont {Galembeck}, \citenamefont {Kapral},\ and\
  \citenamefont {Ozin}}]{ozin:10}%
  \BibitemOpen
  \bibfield  {author} {\bibinfo {author} {\bibfnamefont {L.~F.}\ \bibnamefont
  {Valadares}}, \bibinfo {author} {\bibfnamefont {Y.-G.}\ \bibnamefont {Tao}},
  \bibinfo {author} {\bibfnamefont {N.~S.}\ \bibnamefont {Zacharia}}, \bibinfo
  {author} {\bibfnamefont {V.}~\bibnamefont {Kitaev}}, \bibinfo {author}
  {\bibfnamefont {F.}~\bibnamefont {Galembeck}}, \bibinfo {author}
  {\bibfnamefont {R.}~\bibnamefont {Kapral}}, \ and\ \bibinfo {author}
  {\bibfnamefont {G.~A.}\ \bibnamefont {Ozin}},\ }\href {\doibase
  10.1002/smll.200901976} {\bibfield  {journal} {\bibinfo  {journal} {Small}\
  }\textbf {\bibinfo {volume} {6}},\ \bibinfo {pages} {565} (\bibinfo {year}
  {2010})}\BibitemShut {NoStop}%
\bibitem [{\citenamefont {Sengupta}\ \emph {et~al.}(2014)\citenamefont
  {Sengupta}, \citenamefont {Patra}, \citenamefont {Ortiz-Rivera},
  \citenamefont {Agrawal}, \citenamefont {Shklyaev}, \citenamefont {Dey},
  \citenamefont {C{\'o}rdova-Figueroa}, \citenamefont {Mallouk},\ and\
  \citenamefont {Sen}}]{sengupta:14}%
  \BibitemOpen
  \bibfield  {author} {\bibinfo {author} {\bibfnamefont {S.}~\bibnamefont
  {Sengupta}}, \bibinfo {author} {\bibfnamefont {D.}~\bibnamefont {Patra}},
  \bibinfo {author} {\bibfnamefont {I.}~\bibnamefont {Ortiz-Rivera}}, \bibinfo
  {author} {\bibfnamefont {A.}~\bibnamefont {Agrawal}}, \bibinfo {author}
  {\bibfnamefont {S.}~\bibnamefont {Shklyaev}}, \bibinfo {author}
  {\bibfnamefont {K.~K.}\ \bibnamefont {Dey}}, \bibinfo {author} {\bibfnamefont
  {U.}~\bibnamefont {C{\'o}rdova-Figueroa}}, \bibinfo {author} {\bibfnamefont
  {T.~E.}\ \bibnamefont {Mallouk}}, \ and\ \bibinfo {author} {\bibfnamefont
  {A.}~\bibnamefont {Sen}},\ }\href {\doibase 10.1038/nchem.1895} {\bibfield
  {journal} {\bibinfo  {journal} {Nature Chem.}\ }\textbf {\bibinfo {volume}
  {6}},\ \bibinfo {pages} {415} (\bibinfo {year} {2014})}\BibitemShut {NoStop}%
\bibitem [{\citenamefont {Popescu}\ \emph {et~al.}(2009)\citenamefont
  {Popescu}, \citenamefont {Dietrich},\ and\ \citenamefont
  {Oshanin}}]{popescu:09}%
  \BibitemOpen
  \bibfield  {author} {\bibinfo {author} {\bibfnamefont {M.~N.}\ \bibnamefont
  {Popescu}}, \bibinfo {author} {\bibfnamefont {S.}~\bibnamefont {Dietrich}}, \
  and\ \bibinfo {author} {\bibfnamefont {G.}~\bibnamefont {Oshanin}},\ }\href
  {\doibase 10.1063/1.3133239} {\bibfield  {journal} {\bibinfo  {journal} {J.
  Chem. Phys.}\ }\textbf {\bibinfo {volume} {130}},\ \bibinfo {pages} {194702}
  (\bibinfo {year} {2009})}\BibitemShut {NoStop}%
\bibitem [{\citenamefont {Volpe}\ \emph {et~al.}(2011)\citenamefont {Volpe},
  \citenamefont {Buttinoni}, \citenamefont {Vogt}, \citenamefont
  {K{\"u}mmerer},\ and\ \citenamefont {Bechinger}}]{volpe:11}%
  \BibitemOpen
  \bibfield  {author} {\bibinfo {author} {\bibfnamefont {G.}~\bibnamefont
  {Volpe}}, \bibinfo {author} {\bibfnamefont {I.}~\bibnamefont {Buttinoni}},
  \bibinfo {author} {\bibfnamefont {D.}~\bibnamefont {Vogt}}, \bibinfo {author}
  {\bibfnamefont {H.-J.}\ \bibnamefont {K{\"u}mmerer}}, \ and\ \bibinfo
  {author} {\bibfnamefont {C.}~\bibnamefont {Bechinger}},\ }\href {\doibase
  10.1039/C1SM05960B} {\bibfield  {journal} {\bibinfo  {journal} {Soft Matter}\
  }\textbf {\bibinfo {volume} {7}},\ \bibinfo {pages} {8810} (\bibinfo {year}
  {2011})}\BibitemShut {NoStop}%
\bibitem [{\citenamefont {Uspal}\ \emph {et~al.}(2015)\citenamefont {Uspal},
  \citenamefont {Popescu}, \citenamefont {Dietrich},\ and\ \citenamefont
  {Tasinkevych}}]{uspal:14}%
  \BibitemOpen
  \bibfield  {author} {\bibinfo {author} {\bibfnamefont {W.~E.}\ \bibnamefont
  {Uspal}}, \bibinfo {author} {\bibfnamefont {M.~N.}\ \bibnamefont {Popescu}},
  \bibinfo {author} {\bibfnamefont {S.}~\bibnamefont {Dietrich}}, \ and\
  \bibinfo {author} {\bibfnamefont {M.}~\bibnamefont {Tasinkevych}},\ }\href
  {\doibase 10.1039/c4sm02317j} {\bibfield  {journal} {\bibinfo  {journal}
  {Soft Matter}\ }\textbf {\bibinfo {volume} {11}},\ \bibinfo {pages} {434}
  (\bibinfo {year} {2015})}\BibitemShut {NoStop}%
\bibitem [{\citenamefont {Crowdy}(2013)}]{crowdy:13}%
  \BibitemOpen
  \bibfield  {author} {\bibinfo {author} {\bibfnamefont {D.~G.}\ \bibnamefont
  {Crowdy}},\ }\href {\doibase 10.1017/jfm.2013.510} {\bibfield  {journal}
  {\bibinfo  {journal} {J. Fluid Mech.}\ }\textbf {\bibinfo {volume} {735}},\
  \bibinfo {pages} {473} (\bibinfo {year} {2013})}\BibitemShut {NoStop}%
\bibitem [{\citenamefont {Ghosh}\ \emph {et~al.}(2013)\citenamefont {Ghosh},
  \citenamefont {Misko}, \citenamefont {Marchesoni},\ and\ \citenamefont
  {Nori}}]{ghosh:13}%
  \BibitemOpen
  \bibfield  {author} {\bibinfo {author} {\bibfnamefont {P.}~\bibnamefont
  {Ghosh}}, \bibinfo {author} {\bibfnamefont {V.}~\bibnamefont {Misko}},
  \bibinfo {author} {\bibfnamefont {F.}~\bibnamefont {Marchesoni}}, \ and\
  \bibinfo {author} {\bibfnamefont {F.}~\bibnamefont {Nori}},\ }\href {\doibase
  10.1103/PhysRevLett.110.268301} {\bibfield  {journal} {\bibinfo  {journal}
  {Phys. Rev. Lett.}\ }\textbf {\bibinfo {volume} {110}},\ \bibinfo {pages}
  {268301} (\bibinfo {year} {2013})}\BibitemShut {NoStop}%
\bibitem [{\citenamefont {Ghosh}(2014)}]{ghosh:14}%
  \BibitemOpen
  \bibfield  {author} {\bibinfo {author} {\bibfnamefont {P.~K.}\ \bibnamefont
  {Ghosh}},\ }\href {\doibase 10.1063/1.4892970} {\bibfield  {journal}
  {\bibinfo  {journal} {J. Chem. Phys.}\ }\textbf {\bibinfo {volume} {141}},\
  \bibinfo {pages} {061102} (\bibinfo {year} {2014})}\BibitemShut {NoStop}%
\bibitem [{\citenamefont {Elgeti}\ and\ \citenamefont
  {Gompper}(2009)}]{elgeti:09}%
  \BibitemOpen
  \bibfield  {author} {\bibinfo {author} {\bibfnamefont {J.}~\bibnamefont
  {Elgeti}}\ and\ \bibinfo {author} {\bibfnamefont {G.}~\bibnamefont
  {Gompper}},\ }\href {\doibase 10.1209/0295-5075/85/38002} {\bibfield
  {journal} {\bibinfo  {journal} {EPL}\ }\textbf {\bibinfo {volume} {85}},\
  \bibinfo {pages} {38002} (\bibinfo {year} {2009})}\BibitemShut {NoStop}%
\bibitem [{\citenamefont {Elgeti}\ and\ \citenamefont
  {Gompper}(2013)}]{elgeti:13}%
  \BibitemOpen
  \bibfield  {author} {\bibinfo {author} {\bibfnamefont {J.}~\bibnamefont
  {Elgeti}}\ and\ \bibinfo {author} {\bibfnamefont {G.}~\bibnamefont
  {Gompper}},\ }\href {\doibase 10.1209/0295-5075/101/48003} {\bibfield
  {journal} {\bibinfo  {journal} {EPL}\ }\textbf {\bibinfo {volume} {101}},\
  \bibinfo {pages} {48003} (\bibinfo {year} {2013})}\BibitemShut {NoStop}%
\bibitem [{\citenamefont {Wensink}\ and\ \citenamefont
  {L{\"o}wen}(2008)}]{wensink:08}%
  \BibitemOpen
  \bibfield  {author} {\bibinfo {author} {\bibfnamefont {H.~H.}\ \bibnamefont
  {Wensink}}\ and\ \bibinfo {author} {\bibfnamefont {H.}~\bibnamefont
  {L{\"o}wen}},\ }\href {\doibase 10.1103/PhysRevE.78.031409} {\bibfield
  {journal} {\bibinfo  {journal} {Phys. Rev. E}\ }\textbf {\bibinfo {volume}
  {78}},\ \bibinfo {pages} {031409} (\bibinfo {year} {2008})}\BibitemShut
  {NoStop}%
\bibitem [{\citenamefont {Lauga}\ and\ \citenamefont
  {Powers}(2009)}]{lauga:09}%
  \BibitemOpen
  \bibfield  {author} {\bibinfo {author} {\bibfnamefont {E.}~\bibnamefont
  {Lauga}}\ and\ \bibinfo {author} {\bibfnamefont {T.~R.}\ \bibnamefont
  {Powers}},\ }\href {\doibase 10.1088/0034-4885/72/9/096601} {\bibfield
  {journal} {\bibinfo  {journal} {Rep. Prog. Phys.}\ }\textbf {\bibinfo
  {volume} {72}},\ \bibinfo {pages} {096601} (\bibinfo {year}
  {2009})}\BibitemShut {NoStop}%
\bibitem [{\citenamefont {Spagnolie}\ and\ \citenamefont
  {Lauga}(2012)}]{lauga:12}%
  \BibitemOpen
  \bibfield  {author} {\bibinfo {author} {\bibfnamefont {S.~E.}\ \bibnamefont
  {Spagnolie}}\ and\ \bibinfo {author} {\bibfnamefont {E.}~\bibnamefont
  {Lauga}},\ }\href {\doibase 10.1017/jfm.2012.101} {\bibfield  {journal}
  {\bibinfo  {journal} {J. Fluid Mech.}\ }\textbf {\bibinfo {volume} {700}},\
  \bibinfo {pages} {105} (\bibinfo {year} {2012})}\BibitemShut {NoStop}%
\bibitem [{\citenamefont {Hernandez-Ortiz}\ \emph {et~al.}(2005)\citenamefont
  {Hernandez-Ortiz}, \citenamefont {Stoltz},\ and\ \citenamefont
  {Graham}}]{graham:05}%
  \BibitemOpen
  \bibfield  {author} {\bibinfo {author} {\bibfnamefont {J.~P.}\ \bibnamefont
  {Hernandez-Ortiz}}, \bibinfo {author} {\bibfnamefont {C.~G.}\ \bibnamefont
  {Stoltz}}, \ and\ \bibinfo {author} {\bibfnamefont {M.~D.}\ \bibnamefont
  {Graham}},\ }\href {\doibase 10.1103/PhysRevLett.95.204501} {\bibfield
  {journal} {\bibinfo  {journal} {Phys. Rev. Lett.}\ }\textbf {\bibinfo
  {volume} {95}},\ \bibinfo {pages} {204501} (\bibinfo {year}
  {2005})}\BibitemShut {NoStop}%
\bibitem [{\citenamefont {Hernandez-Ortiz}\ \emph {et~al.}(2009)\citenamefont
  {Hernandez-Ortiz}, \citenamefont {Underhill},\ and\ \citenamefont
  {Graham}}]{graham:09}%
  \BibitemOpen
  \bibfield  {author} {\bibinfo {author} {\bibfnamefont {J.~P.}\ \bibnamefont
  {Hernandez-Ortiz}}, \bibinfo {author} {\bibfnamefont {P.~T.}\ \bibnamefont
  {Underhill}}, \ and\ \bibinfo {author} {\bibfnamefont {M.~D.}\ \bibnamefont
  {Graham}},\ }\href {\doibase 10.1088/0953-8984/21/20/204107} {\bibfield
  {journal} {\bibinfo  {journal} {J. Phys.: Condens. Matter}\ }\textbf
  {\bibinfo {volume} {21}},\ \bibinfo {pages} {204107} (\bibinfo {year}
  {2009})}\BibitemShut {NoStop}%
\bibitem [{\citenamefont {Z{\"o}ttl}\ and\ \citenamefont
  {Stark}(2014)}]{zoettl:14}%
  \BibitemOpen
  \bibfield  {author} {\bibinfo {author} {\bibfnamefont {A.}~\bibnamefont
  {Z{\"o}ttl}}\ and\ \bibinfo {author} {\bibfnamefont {H.}~\bibnamefont
  {Stark}},\ }\href {\doibase 10.1103/PhysRevLett.112.118101} {\bibfield
  {journal} {\bibinfo  {journal} {Phys. Rev. Lett.}\ }\textbf {\bibinfo
  {volume} {112}},\ \bibinfo {pages} {118101} (\bibinfo {year}
  {2014})}\BibitemShut {NoStop}%
\bibitem [{\citenamefont {Pavlick}\ \emph {et~al.}(2013)\citenamefont
  {Pavlick}, \citenamefont {Dey}, \citenamefont {Sirjoosingh}, \citenamefont
  {Benesi},\ and\ \citenamefont {Sen}}]{pavlick:13}%
  \BibitemOpen
  \bibfield  {author} {\bibinfo {author} {\bibfnamefont {R.~A.}\ \bibnamefont
  {Pavlick}}, \bibinfo {author} {\bibfnamefont {K.~K.}\ \bibnamefont {Dey}},
  \bibinfo {author} {\bibfnamefont {A.}~\bibnamefont {Sirjoosingh}}, \bibinfo
  {author} {\bibfnamefont {A.}~\bibnamefont {Benesi}}, \ and\ \bibinfo {author}
  {\bibfnamefont {A.}~\bibnamefont {Sen}},\ }\href {\doibase
  10.1039/C2NR32518G} {\bibfield  {journal} {\bibinfo  {journal} {Nanoscale}\
  }\textbf {\bibinfo {volume} {5}},\ \bibinfo {pages} {1301} (\bibinfo {year}
  {2013})}\BibitemShut {NoStop}%
\bibitem [{\citenamefont {Muddana}\ \emph {et~al.}(2010)\citenamefont
  {Muddana}, \citenamefont {Sengupta}, \citenamefont {Mallouk}, \citenamefont
  {Sen},\ and\ \citenamefont {Butler}}]{muddana:10}%
  \BibitemOpen
  \bibfield  {author} {\bibinfo {author} {\bibfnamefont {H.~S.}\ \bibnamefont
  {Muddana}}, \bibinfo {author} {\bibfnamefont {S.}~\bibnamefont {Sengupta}},
  \bibinfo {author} {\bibfnamefont {T.~E.}\ \bibnamefont {Mallouk}}, \bibinfo
  {author} {\bibfnamefont {A.}~\bibnamefont {Sen}}, \ and\ \bibinfo {author}
  {\bibfnamefont {P.~J.}\ \bibnamefont {Butler}},\ }\href {\doibase
  10.1021/ja908773a} {\bibfield  {journal} {\bibinfo  {journal} {J. Am. Chem.
  Soc.}\ }\textbf {\bibinfo {volume} {132}},\ \bibinfo {pages} {2110} (\bibinfo
  {year} {2010})}\BibitemShut {NoStop}%
\bibitem [{\citenamefont {Sengupta}\ \emph {et~al.}(2013)\citenamefont
  {Sengupta}, \citenamefont {Dey}, \citenamefont {Muddana}, \citenamefont
  {Tabouillot}, \citenamefont {Ibele}, \citenamefont {Butler},\ and\
  \citenamefont {Sen}}]{sengupta:13}%
  \BibitemOpen
  \bibfield  {author} {\bibinfo {author} {\bibfnamefont {S.}~\bibnamefont
  {Sengupta}}, \bibinfo {author} {\bibfnamefont {K.~K.}\ \bibnamefont {Dey}},
  \bibinfo {author} {\bibfnamefont {H.~S.}\ \bibnamefont {Muddana}}, \bibinfo
  {author} {\bibfnamefont {T.}~\bibnamefont {Tabouillot}}, \bibinfo {author}
  {\bibfnamefont {M.~E.}\ \bibnamefont {Ibele}}, \bibinfo {author}
  {\bibfnamefont {P.~J.}\ \bibnamefont {Butler}}, \ and\ \bibinfo {author}
  {\bibfnamefont {A.}~\bibnamefont {Sen}},\ }\href {\doibase 10.1021/ja3091615}
  {\bibfield  {journal} {\bibinfo  {journal} {J. Am. Chem. Soc.}\ }\textbf
  {\bibinfo {volume} {135}},\ \bibinfo {pages} {1406} (\bibinfo {year}
  {2013})}\BibitemShut {NoStop}%
\bibitem [{\citenamefont {Lee}\ \emph {et~al.}(2014)\citenamefont {Lee},
  \citenamefont {Alarc{\'o}n-Correa}, \citenamefont {Miksch}, \citenamefont
  {Hahn}, \citenamefont {Gibbs},\ and\ \citenamefont {Fischer}}]{lee:14}%
  \BibitemOpen
  \bibfield  {author} {\bibinfo {author} {\bibfnamefont {T.-C.}\ \bibnamefont
  {Lee}}, \bibinfo {author} {\bibfnamefont {M.}~\bibnamefont
  {Alarc{\'o}n-Correa}}, \bibinfo {author} {\bibfnamefont {C.}~\bibnamefont
  {Miksch}}, \bibinfo {author} {\bibfnamefont {K.}~\bibnamefont {Hahn}},
  \bibinfo {author} {\bibfnamefont {J.~G.}\ \bibnamefont {Gibbs}}, \ and\
  \bibinfo {author} {\bibfnamefont {P.}~\bibnamefont {Fischer}},\ }\href
  {\doibase 10.1021/nl500068n} {\bibfield  {journal} {\bibinfo  {journal} {Nano
  Lett.}\ }\textbf {\bibinfo {volume} {14}},\ \bibinfo {pages} {2407} (\bibinfo
  {year} {2014})}\BibitemShut {NoStop}%
\bibitem [{\citenamefont {G{\'a}sp{\'a}r}(2014)}]{gaspar:14}%
  \BibitemOpen
  \bibfield  {author} {\bibinfo {author} {\bibfnamefont {S.}~\bibnamefont
  {G{\'a}sp{\'a}r}},\ }\href {\doibase 10.1039/C4NR01760A} {\bibfield
  {journal} {\bibinfo  {journal} {Nanoscale}\ }\textbf {\bibinfo {volume}
  {6}},\ \bibinfo {pages} {7757} (\bibinfo {year} {2014})}\BibitemShut
  {NoStop}%
\bibitem [{\citenamefont {Colberg}\ and\ \citenamefont
  {Kapral}(2014)}]{colberg:14}%
  \BibitemOpen
  \bibfield  {author} {\bibinfo {author} {\bibfnamefont {P.~H.}\ \bibnamefont
  {Colberg}}\ and\ \bibinfo {author} {\bibfnamefont {R.}~\bibnamefont
  {Kapral}},\ }\href {\doibase 10.1209/0295-5075/106/30004} {\bibfield
  {journal} {\bibinfo  {journal} {EPL}\ }\textbf {\bibinfo {volume} {106}},\
  \bibinfo {pages} {30004} (\bibinfo {year} {2014})}\BibitemShut {NoStop}%
\bibitem [{\citenamefont {Anderson}\ and\ \citenamefont
  {Prieve}(1984)}]{anderson:84}%
  \BibitemOpen
  \bibfield  {author} {\bibinfo {author} {\bibfnamefont {J.~L.}\ \bibnamefont
  {Anderson}}\ and\ \bibinfo {author} {\bibfnamefont {D.~C.}\ \bibnamefont
  {Prieve}},\ }\href {\doibase 10.1080/03602548408068407} {\bibfield  {journal}
  {\bibinfo  {journal} {Sep. Purif. Rev.}\ }\textbf {\bibinfo {volume} {13}},\
  \bibinfo {pages} {67} (\bibinfo {year} {1984})}\BibitemShut {NoStop}%
\bibitem [{\citenamefont {Anderson}(1989)}]{anderson:89}%
  \BibitemOpen
  \bibfield  {author} {\bibinfo {author} {\bibfnamefont {J.~L.}\ \bibnamefont
  {Anderson}},\ }\href {\doibase 10.1146/annurev.fl.21.010189.000425}
  {\bibfield  {journal} {\bibinfo  {journal} {Ann. Rev. Fluid Mech.}\ }\textbf
  {\bibinfo {volume} {21}},\ \bibinfo {pages} {61} (\bibinfo {year}
  {1989})}\BibitemShut {NoStop}%
\bibitem [{\citenamefont {Golestanian}\ \emph {et~al.}(2005)\citenamefont
  {Golestanian}, \citenamefont {Liverpool},\ and\ \citenamefont
  {Ajdari}}]{goles:05}%
  \BibitemOpen
  \bibfield  {author} {\bibinfo {author} {\bibfnamefont {R.}~\bibnamefont
  {Golestanian}}, \bibinfo {author} {\bibfnamefont {T.~B.}\ \bibnamefont
  {Liverpool}}, \ and\ \bibinfo {author} {\bibfnamefont {A.}~\bibnamefont
  {Ajdari}},\ }\href {\doibase 10.1103/PhysRevLett.94.220801} {\bibfield
  {journal} {\bibinfo  {journal} {Phys. Rev. Lett.}\ }\textbf {\bibinfo
  {volume} {94}},\ \bibinfo {pages} {220801} (\bibinfo {year}
  {2005})}\BibitemShut {NoStop}%
\bibitem [{\citenamefont {R{\"u}ckner}\ and\ \citenamefont
  {Kapral}(2007)}]{kapral:07}%
  \BibitemOpen
  \bibfield  {author} {\bibinfo {author} {\bibfnamefont {G.}~\bibnamefont
  {R{\"u}ckner}}\ and\ \bibinfo {author} {\bibfnamefont {R.}~\bibnamefont
  {Kapral}},\ }\href {\doibase 10.1103/PhysRevLett.98.150603} {\bibfield
  {journal} {\bibinfo  {journal} {Phys. Rev. Lett.}\ }\textbf {\bibinfo
  {volume} {98}},\ \bibinfo {pages} {150603} (\bibinfo {year}
  {2007})}\BibitemShut {NoStop}%
\bibitem [{\citenamefont {Tao}\ and\ \citenamefont {Kapral}(2008)}]{yuguo:08}%
  \BibitemOpen
  \bibfield  {author} {\bibinfo {author} {\bibfnamefont {Y.-G.}\ \bibnamefont
  {Tao}}\ and\ \bibinfo {author} {\bibfnamefont {R.}~\bibnamefont {Kapral}},\
  }\href {\doibase 10.1063/1.2908078} {\bibfield  {journal} {\bibinfo
  {journal} {J. Chem. Phys.}\ }\textbf {\bibinfo {volume} {128}},\ \bibinfo
  {pages} {164518} (\bibinfo {year} {2008})}\BibitemShut {NoStop}%
\bibitem [{\citenamefont {Colberg}\ \emph {et~al.}(2014)\citenamefont
  {Colberg}, \citenamefont {Reigh}, \citenamefont {Robertson},\ and\
  \citenamefont {Kapral}}]{kapral:14}%
  \BibitemOpen
  \bibfield  {author} {\bibinfo {author} {\bibfnamefont {P.~H.}\ \bibnamefont
  {Colberg}}, \bibinfo {author} {\bibfnamefont {S.~Y.}\ \bibnamefont {Reigh}},
  \bibinfo {author} {\bibfnamefont {B.}~\bibnamefont {Robertson}}, \ and\
  \bibinfo {author} {\bibfnamefont {R.}~\bibnamefont {Kapral}},\ }\href
  {\doibase 10.1021/ar5002582} {\bibfield  {journal} {\bibinfo  {journal} {Acc.
  Chem. Res.}\ }\textbf {\bibinfo {volume} {47}},\ \bibinfo {pages} {3504}
  (\bibinfo {year} {2014})}\BibitemShut {NoStop}%
\bibitem [{\citenamefont {Yang}\ \emph {et~al.}(2014)\citenamefont {Yang},
  \citenamefont {Wysockia},\ and\ \citenamefont {Ripoll}}]{yang:14}%
  \BibitemOpen
  \bibfield  {author} {\bibinfo {author} {\bibfnamefont {M.}~\bibnamefont
  {Yang}}, \bibinfo {author} {\bibfnamefont {A.}~\bibnamefont {Wysockia}}, \
  and\ \bibinfo {author} {\bibfnamefont {M.}~\bibnamefont {Ripoll}},\ }\href
  {\doibase 10.1039/c4sm00621f} {\bibfield  {journal} {\bibinfo  {journal}
  {Soft Matter}\ }\textbf {\bibinfo {volume} {10}},\ \bibinfo {pages} {6208}
  (\bibinfo {year} {2014})}\BibitemShut {NoStop}%
\bibitem [{\citenamefont {Popescu}\ \emph {et~al.}(2011)\citenamefont
  {Popescu}, \citenamefont {Tasinkevych},\ and\ \citenamefont
  {Dietrich}}]{popescu:11}%
  \BibitemOpen
  \bibfield  {author} {\bibinfo {author} {\bibfnamefont {M.~N.}\ \bibnamefont
  {Popescu}}, \bibinfo {author} {\bibfnamefont {M.}~\bibnamefont
  {Tasinkevych}}, \ and\ \bibinfo {author} {\bibfnamefont {S.}~\bibnamefont
  {Dietrich}},\ }\href {\doibase 10.1209/0295-5075/95/28004} {\bibfield
  {journal} {\bibinfo  {journal} {EPL}\ }\textbf {\bibinfo {volume} {95}},\
  \bibinfo {pages} {28004} (\bibinfo {year} {2011})}\BibitemShut {NoStop}%
\bibitem [{\citenamefont {Reigh}\ and\ \citenamefont
  {Kapral}(2015)}]{reigh:15}%
  \BibitemOpen
  \bibfield  {author} {\bibinfo {author} {\bibfnamefont {S.~Y.}\ \bibnamefont
  {Reigh}}\ and\ \bibinfo {author} {\bibfnamefont {R.}~\bibnamefont {Kapral}},\
  }\href {\doibase 10.1039/C4SM02857K} {\bibfield  {journal} {\bibinfo
  {journal} {Soft Matter}\ }\textbf {\bibinfo {volume} {11}},\ \bibinfo {pages}
  {3149} (\bibinfo {year} {2015})}\BibitemShut {NoStop}%
\bibitem [{Note1()}]{Note1}%
  \BibitemOpen
  \bibinfo {note} {The simulations were run on GPUs using a massively parallel
  code written in OpenCL~C and Lua, which is distributed under a free software
  license at \protect \url {http://colberg.org/angstrom-dimer}.}\BibitemShut
  {Stop}%
\bibitem [{Note2()}]{Note2}%
  \BibitemOpen
  \bibinfo {note} {In applications to experimental systems the specific
  properties of the intermolecular potentials should be taken into
  account.}\BibitemShut {Stop}%
\bibitem [{\citenamefont {Rahman}(1964)}]{rahman:64}%
  \BibitemOpen
  \bibfield  {author} {\bibinfo {author} {\bibfnamefont {A.}~\bibnamefont
  {Rahman}},\ }\href {\doibase 10.1103/PhysRev.136.A405} {\bibfield  {journal}
  {\bibinfo  {journal} {Phys. Rev.}\ }\textbf {\bibinfo {volume} {136}},\
  \bibinfo {pages} {A405} (\bibinfo {year} {1964})}\BibitemShut {NoStop}%
\bibitem [{\citenamefont {Lekkerkerker}\ and\ \citenamefont
  {Tuinier}(2011)}]{lekkerkerker:11}%
  \BibitemOpen
  \bibfield  {author} {\bibinfo {author} {\bibfnamefont {H.~N.}\ \bibnamefont
  {Lekkerkerker}}\ and\ \bibinfo {author} {\bibfnamefont {R.}~\bibnamefont
  {Tuinier}},\ }\href {\doibase 10.1007/978-94-007-1223-2} {\emph {\bibinfo
  {title} {Colloids and the Depletion Interaction}}}\ (\bibinfo  {publisher}
  {Springer Netherlands},\ \bibinfo {year} {2011})\BibitemShut {NoStop}%
\bibitem [{Note3()}]{Note3}%
  \BibitemOpen
  \bibinfo {note} {Ensemble averages were carried out over 30 realisations of
  each $10^8$ integration steps for a given system $L_Z$, $\epsilon _\protect
  \text {NB}$, and $\epsilon _{\protect \text {w}S}$.}\BibitemShut {Stop}%
\bibitem [{\citenamefont {Bickel}(2007)}]{bickel:07}%
  \BibitemOpen
  \bibfield  {author} {\bibinfo {author} {\bibfnamefont {T.}~\bibnamefont
  {Bickel}},\ }\href {\doibase 10.1016/j.physa.2006.11.008} {\bibfield
  {journal} {\bibinfo  {journal} {Physica A}\ }\textbf {\bibinfo {volume}
  {377}},\ \bibinfo {pages} {24} (\bibinfo {year} {2007})}\BibitemShut
  {NoStop}%
\bibitem [{Note4()}]{Note4}%
  \BibitemOpen
  \bibinfo {note} {A comparable scaling behavior has been derived for an
  unrelated system with superdiffusive transport, where the first-passage time
  is found to follow a power law with exponent $\protect \frac {4}{3}$.~\cite
  {redner:90}}\BibitemShut {NoStop}%
\bibitem [{\citenamefont {Bouchaud}\ \emph {et~al.}(1990)\citenamefont
  {Bouchaud}, \citenamefont {Georges}, \citenamefont {Koplik}, \citenamefont
  {Provata},\ and\ \citenamefont {Redner}}]{redner:90}%
  \BibitemOpen
  \bibfield  {author} {\bibinfo {author} {\bibfnamefont {J.-P.}\ \bibnamefont
  {Bouchaud}}, \bibinfo {author} {\bibfnamefont {A.}~\bibnamefont {Georges}},
  \bibinfo {author} {\bibfnamefont {J.}~\bibnamefont {Koplik}}, \bibinfo
  {author} {\bibfnamefont {A.}~\bibnamefont {Provata}}, \ and\ \bibinfo
  {author} {\bibfnamefont {S.}~\bibnamefont {Redner}},\ }\href {\doibase
  10.1103/PhysRevLett.64.2503} {\bibfield  {journal} {\bibinfo  {journal}
  {Phys. Rev. Lett.}\ }\textbf {\bibinfo {volume} {64}},\ \bibinfo {pages}
  {2503} (\bibinfo {year} {1990})}\BibitemShut {NoStop}%
\end{thebibliography}%

\end{document}